\documentclass[aps,prb, twocolumn, superscriptaddress, noeprint]{revtex4-2}

\usepackage{amsmath,amssymb,graphicx,dcolumn,bm,hyperref,float}

\hypersetup{ colorlinks=true, linkcolor=red, filecolor=red,  urlcolor=red, citecolor = red}

\renewcommand{\vec}[1]{\boldsymbol{#1}}
\newcommand{\RNum}[1]{\uppercase\expandafter{\romannumeral #1\relax}}

\def \beq {\begin{eqnarray}}
\def \eeq {\end{eqnarray}}

\begin{document}

\title{Emergent $\mathbb{Z}_2$ symmetry near a CDW multicritical point}
\author{Steven A. Kivelson}
\affiliation{Department of Physics, Stanford University, Stanford, CA 94305, USA}
\affiliation{Rudolf Peierls Centre for Theoretical Physics, University of Oxford, Oxford OX1 3PU, United Kingdom}
\affiliation{Stanford Institute for Materials and Energy Sciences, SLAC National Accelerator Laboratory,
2575 Sand Hill Road, Menlo Park, California 94025, USA}

\author{   Akshat Pandey} 
\affiliation{Department of Physics, Stanford University, Stanford, CA 94305, USA}
\affiliation{Rudolf Peierls Centre for Theoretical Physics, University of Oxford, Oxford OX1 3PU, United Kingdom}

\author{Anisha G. Singh}
\affiliation{Stanford Institute for Materials and Energy Sciences, SLAC National Accelerator Laboratory,
2575 Sand Hill Road, Menlo Park, California 94025, USA}
\affiliation{Department of Applied Physics, Stanford University, Stanford, CA 94305, USA}

\author{Aharon Kapitulnik}
\affiliation{Department of Physics, Stanford University, Stanford, CA 94305, USA}
\affiliation{Stanford Institute for Materials and Energy Sciences, SLAC National Accelerator Laboratory,
2575 Sand Hill Road, Menlo Park, California 94025, USA}
\affiliation{Department of Applied Physics, Stanford University, Stanford, CA 94305, USA}
\author{Ian R. Fisher}
\affiliation{Stanford Institute for Materials and Energy Sciences, SLAC National Accelerator Laboratory,
2575 Sand Hill Road, Menlo Park, California 94025, USA}
\affiliation{Department of Applied Physics, Stanford University, Stanford, CA 94305, USA}

\date{\today}

\begin{abstract}
We consider the critical behavior associated with incommensurate unidirectional charge-density-wave ordering in a weakly orthorhombic system subject to uniaxial strain as an experimentally significant example of $U(1)\times U(1)$ multicriticality.  We show that, depending on microscopic details, the phase diagram can have qualitatively different structures which can involve a vestigial metanematic critical point, a pair of tricritical points, a decoupled tetracritical point, or (at least at mean-field level) a bicritical point.  We analyze the emergent symmetries in the critical regime and find that these can --- at least in some cases --- involve an emergent $\mathbb{Z}_2$ order parameter symmetry.
\end{abstract}
\maketitle

From an electronic structure perspective, ErTe$_3$ is very nearly tetragonal, but from a structural perspective, due to the presence of a glide plane, it is intrinsically orthorhombic~\cite{RTe3}.  
Below a well-characterized transition temperature, $T_c$, it exhibits unidirectional incommensurate charge-density-wave (CDW) order with ordering wave vector along the orthorhombic $c$ axis. In the presence of an in-plane unidirectional applied stress $s$ in excess of a modest critical value, $s^*$ (or, more generally, anisotropic in-plane strain greater than some critical value), the CDW ordering wave vector rotates by $\pi/2$ to lie along the $a$ axis ~\cite{Joshexpt,Z2experiment,Z2experiment2}. Moreover, for temperatures slightly above $T_c$, the nematic susceptibility (as inferred from the elastoresistance) as a function of strain is strongly peaked at $s=s^*$~\cite{Z2experiment}.  These observations motivate us to reconsider the possible multicritical phase diagrams relevant to this general situation in which there are two distinct $U(1)$ order parameters (associated with breaking of translational symmetry in the two directions).

From a symmetry perspective, the two CDW states found for ErTe$_3$ are fundamentally distinct; this is an inescapable consequence of the orthorhombic crystal structure. For instance, while the magnitudes of the ordering wave vectors in the two directions are similar, they are not identical ~\cite{Z2experiment}. Similarly, the magnitude of the slope of $T_c$ with respect to $s$ is different for the two states ~\cite{Z2experiment}. Nonetheless,
we identify circumstances when, in a sense that we will define precisely, there is an  emergent $\mathbb{Z}_2$ symmetry under exchange of the two order parameters that characterizes the critical regime  --- i.e., the system behaves as if it were truly tetragonal at a critical value of the stress, $s=s^\star$, and for $T$ close to an appropriate critical point. 

Specifically, we consider a Landau-Ginzburg-Wilson effective field theory with two complex scalar order parameters, $\phi_1$ and $\phi_2$, corresponding to the two components of the CDW order.  We will consider the solution of this problem in the multicritical regime first in the context of Landau mean-field theory. Then, more qualitatively (and conjecturally) we will discuss the true three-dimensional critical fluctuations. 

For simplicity, we will start by treating the  case in which the system is actually tetragonal, in which case, for vanishing applied stress ($s=0$), there is a set of discrete (point-group) symmetries which interchange $\phi_1 \ \leftrightarrow \ \phi_2$ and at the same time interchange coordinates, $x\ \leftrightarrow\ y$~\footnote{For simplicity, we will refer to this as a $\mathbb{Z}_2$ symmetry reflecting its action on the order parameters, but since the true symmetries at play are spatial symmetries, this nomenclature should not be taken literally.}.   We will then analyze the effects of including small terms consistent with an orthorhombic point group symmetry. In Figs.~\ref{fig:MFPhaseDiagrams} and \ref{fig:orthoMFPhaseDiagrams} we show the possible structures of the phase diagram that come from a saddle-point (Landau mean-field) solution of the model in various ranges of couplings for the tetragonal system and the weakly orthorhombic system respectively. 

The effective field theories from which these results are derived are defined in Sec.~\ref{sec:eft}.  The  mean-field treatment and the considerations concerning the true asymptotic critical phenomena are  discussed in Secs.~\ref{sec:meanfield} and \ref{sec:fluctuations}, respectively.  In Sec.~\ref{sec:discussion}, the results are discussed with emphasis on the existence (or not) of additional emergent symmetries in the asymptotic near-critical regime.

\section{The Effective Field Theory}\label{sec:eft}

The effective (Landau-Ginzburg-Wilson) field theory in the neighborhood of a multicritical point for two unidirectional incommensurate CDW orders is described by a classical effective Hamiltonian density
\begin{equation}
{\cal H}[\phi_1,\phi_2]= 
 H+\tilde H=V+K  + \tilde V+\tilde K 
\end{equation}
where $H = V+K$ contains all the terms that are consistent with the point group symmetry of a tetragonal system ($D_{4h}$) and an assumed emergent translational symmetry in all three directions (${\mathbb R}^3$), while $\tilde H = \tilde V + \tilde K$ contains the additional terms that are allowed when (either due to applied strain or the intrinsic crystal structure) the point-group symmetry is reduced to that of an orthorhombic system ($D_{2h}$)~\footnote{In unstrained Er$Te_3$, the relevant crystal symmetry group is actually nonsymmorphic, but this does not change the symmetry analysis presented here.}. In this treatment, because the CDW is assumed to be incommensurate, translations also act on the order parameter fields, such that under translation by $\vec a$, the order parameters transform as $\phi_1(\vec r)\to \phi_1(\vec r + \vec a) e^{iQ_xa_x}$ and $\phi_2(\vec r)\to \phi_2(\vec r + \vec a) e^{iQ_ya_y}$ (where $Q_{x,y}$ are the CDW ordering wave vectors in the $x$ and $y$ directions respectively), while any point group element that exchanges $x$ and $y$ also exchanges $\phi_1$ and $\phi_2$.
In the field-theoretic description, 
the translational symmetries in the $x$ and $y$ directions discussed above are manifested as internal $U(1)$ symmetries associated with the order parameter fields, $\phi_1$ and $\phi_2$, while the $\mathbb{Z}_2$   symmetry under exchange of $\phi_1$ and $\phi_2$ requires the simultaneous exchange of $x$ and $y$.

Here $V+\tilde V$  is the effective potential, which we will express as a polynomial in powers of $\phi_1$ and $\phi_2$, keeping explicitly terms only to the  lowest necessary order, while $K+\tilde K$ are the lowest-order gradient terms. 
Explicit expressions for these various terms in the effective Hamiltonian follow:
 \begin{eqnarray}
 \label{Hamiltonian}
&&V =  \frac {\mu}2 \left[ |\phi_1|^2 + |\phi_2|^2\right] + \frac u4\left[ |\phi_1|^2 + |\phi_2|^2\right]^2 
+ \frac \gamma 2 |\phi_1|^2  |\phi_2|^2, \nonumber \\
&&K= \frac{\kappa_L}{2}\left[ |\partial_x\phi_1|^2 + |\partial_y\phi_2|^2\right]+ \frac{\kappa_T}{2}\left[ |\partial_y\phi_1|^2 + |\partial_x\phi_2|^2\right]\nonumber \\
&&\ \ \ \ \ + \frac{\kappa_\perp}{2}\left[ |\partial_z\phi_1|^2 + |\partial_z\phi_2|^2\right] ,
 \label{LGW}
\end{eqnarray}
\begin{eqnarray}
&&\tilde V= \frac {\tilde\mu}2 \left[ |\phi_1|^2 - |\phi_2|^2\right] + \frac {\tilde u}4\left[ |\phi_1|^4 - |\phi_2|^4\right],
\nonumber \\
&&\tilde K = \frac{\tilde\kappa_L}{2}\left[ |\partial_x\phi_1|^2 - |\partial_y\phi_2|^2\right]+ \frac{\tilde \kappa_T}{2}\left[ |\partial_y\phi_1|^2 - |\partial_x\phi_2|^2\right] \nonumber \\
&&\ \ \ \ \ + \frac{\tilde \kappa_\perp}{2}\left[ |\partial_z\phi_1|^2 - |\partial_z\phi_2|^2\right] .
\end{eqnarray}
In most cases, we assume $u > \sqrt{\tilde u^2 + \gamma^2}$, $\kappa_L > |\tilde \kappa_L|$, $\kappa_T > |\tilde \kappa_T|$, and $\kappa_\perp > |\tilde \kappa_\perp|$.  However, when we consider the tricritical phase diagrams in Figs.~\ref{fig:MFPhaseDiagrams}e and \ref{fig:orthoMFPhaseDiagrams}e, we will consider $u$ slightly negative, in which case terms of order at least $\phi^6$ are required  for thermodynamic stability.  Similarly, we shall see that, for the special case $\gamma=0$, higher order terms are necessary (at mean-field level) to fully determine the phase diagram;  where such higher-order terms enter our discussion we will introduce them explicitly.

In general, all the coefficients appearing in these expressions are functions of the temperature, $T$, and if we apply stress to the system,  of the stress tensor, $s_{ab}$.  In an otherwise tetragonal system, all the coefficients with a tilde vanish in the absence of shear strain, and are  odd functions of the   shear stress, $s_{B_{1g}}=s_{xx}-s_{yy}$, while  the coefficients that do not carry tildes are even functions of all other components~\footnote{In Ref. \cite{Z2experiment}, the system is studied at fixed strain, $\epsilon_{ab}$.  In many circumstances the strain is proportional to the stress, so this distinction is of little importance, but where first-order transitions arise, the phase diagrams at fixed strain exhibits regions of macroscopic two-phase coexistence rather than a single first-order phase boundary.}.

\section{Mean-field analysis}\label{sec:meanfield}
In a mean-field analysis, we look for field configurations that minimize ${\cal H}[\phi_1,\phi_2]$.  As $K+\tilde K$ is minimized by any uniform field configuration, this means finding the minima of $V+\tilde V$.

\subsection{The $\mathbb{Z}_2$ symmetric case}\label{subsec:tetrameanfield}
\begin{figure*}
    \centering
    \includegraphics[width=\linewidth]{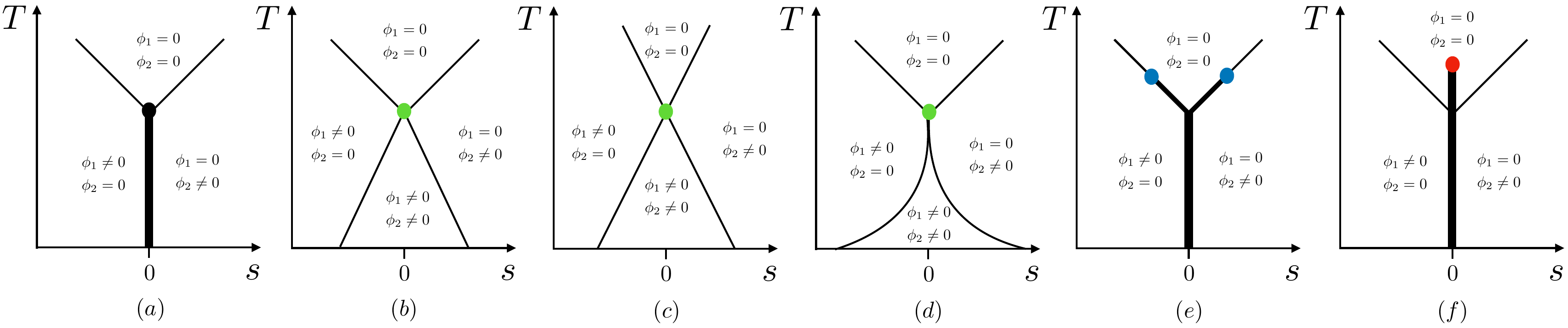}
    \caption{Mean-field phase diagrams in the $\mathbb{Z}_2$-symmetric (tetragonal) cases discussed in Sec.~\ref{subsec:tetrameanfield}:  (a) bicritical point, (b) tetracritical point, (c)  decoupled tetracritical point,  (d) $O(4)$-symmetric tetracritical point, (e) tricritical points, and (f) vestigial nematic. Thin and thick black lines denote, respectively, continuous and first-order transitions.  }
    \label{fig:MFPhaseDiagrams}
\end{figure*}

The mean-field phase diagram in the stress-temperature plane for  the simple case in which the underlying problem is  tetragonal  
is shown in Fig.~\ref{fig:MFPhaseDiagrams} --- also see, e.g., Refs.~\cite{LiuFisher, NelsonKosterlitzFisher, KosterlitzNelsonFisher}.  In making this figure, we have taken $\mu$ and $\tilde \mu$ to be linear in $T$ and $s$ respectively, $\mu=\mu_0[T-T_0]$ and $\tilde\mu=\tilde \mu_0 s$, and have ignored the $T$ and $s$ dependence of all other parameters.  In particular, this means we have set $\tilde u=0$.
Figs.~\ref{fig:MFPhaseDiagrams}a, b, and c follow directly from minimizing $V$ in Eq.~\eqref{LGW}, while Figs.~\ref{fig:MFPhaseDiagrams}d, e, and f involve consideration of additional terms.

The various cases are as follows:
\begin{itemize}
\item  {\bf Fig.~\ref{fig:MFPhaseDiagrams}a:}  For $u>0$ and $\gamma >0$ there is a bicritical point at $\tilde \mu=\mu=0$.  There is a first-order line between two unidirectional phases that occurs along the  half line $\tilde \mu=0$ with $\mu < 0$.  The phase boundaries between the unidirectional ordered phases and the disordered high-temperature phase 
correspond to $\tilde \mu =\pm \mu$ with $\mu\geq 0$.

\item  {\bf Fig.~\ref{fig:MFPhaseDiagrams}b:} For $u>0$ and $-2u< \gamma < 0$ 
there is a tetracritical point at $\mu=\tilde \mu =0$.  For $\tilde \mu >0$, there is a transition from a disordered state for $\mu > |\tilde \mu|$ to a state in which $|\phi_2|>|\phi_1|=0$ at $\mu< |\tilde \mu|$. There is then a further transition to a state with coexisting bidirectional order (but with $\phi_2$ dominant) at $\mu = -|\tilde \mu| (2u-|\gamma|)/|\gamma| $.  For $\tilde \mu <0$, the phase boundaries follow the corresponding lines, but with the roles of $\phi_1$ and $\phi_2$ interchanged.  

\item  {\bf Fig.~\ref{fig:MFPhaseDiagrams}c:} For  $u>0$ and $\gamma=-u$,  
the system has the special feature of exhibiting a ``decoupled'' tetracritical point in which the two phase boundaries pass through each other without changing slope.  At mean-field level, this corresponds to a third degree of fine-tuning, requiring tuning three parameters, $\mu$, $\tilde \mu$, and $\gamma$.

\item  {\bf Fig.~\ref{fig:MFPhaseDiagrams}d:} For  $u>0$ and $\gamma=0$, $V$  exhibits an enhanced $O(4)$ symmetry.  Here there is no coexistence region.  However, there is no latent heat associated with crossing the phase boundary at  $\mu <0$  between the two phases (as $\tilde \mu$ changes sign) so the transition is not conventionally first order despite the fact that the order parameter changes direction abruptly.  Indeed, as is discussed in Ref. \cite{OurCDW}, in this case, the nature of the phase diagram below the multicritical point depends on higher order terms in powers of $\phi_j$.  Specifically, the tetracritical phase diagram shown in Fig.~\ref{fig:MFPhaseDiagrams}d  arises if we include next order terms
\begin{eqnarray}
&&V \to V + \frac {u_6} 6 \left[|\phi_1|^2+|\phi_2|^2\right]^3 \nonumber\\
&& \ \ \ \ \ + \frac {\gamma_6} 3 \left[|\phi_1|^2+|\phi_2|^2\right] |\phi_1|^2|\phi_2|^2
\label{phi6}
\end{eqnarray}
with  $\gamma_6 <0$ (and with $u_6>|\gamma_6|/2$  to preserve stability).  Again, at mean-field level, this form of multicriticality involves an additional degree of fine-tuning relative to a generic tetracritical point.

\item {\bf Fig.~\ref{fig:MFPhaseDiagrams}e:} For $u<0$ the transition at zero stress is necessarily first order.  Generically the expansion in powers of $\phi$ is not justified at a first-order transition unless the transition is only weakly first order, as happens near a tricritical (or bicritical) point~\cite{DomanyTricritical}.  Thus, in Fig.~\ref{fig:MFPhaseDiagrams}e, we have shown the phase diagram that results from a case where $u$ changes sign in the vicinity of a putative bicritical point.  Specifically, we have included an implicit (linear) $T$ dependence of $u$ taking the form  $u=-u^* +u' \mu$,  so that $u<0$ but of small magnitude as $\mu \to 0$, but $u > 0$ for $\mu>\mu^*\equiv u^*/u'$.  To ensure the stability of the free energy, we are forced to include higher-order terms in the effective potential as in Eq.~\ref{phi6}.  To be concrete, we have taken $\gamma>0$.   Now, in the vicinity of the point at which, for $u>0$, there would have been a bicritical point, we instead find a pair of tricritical points at $\mu=\mu^*$ and $\tilde \mu = \pm \mu^\star$. 

\item {\bf In Fig.~\ref{fig:MFPhaseDiagrams}f} we show a phase diagram that cannot be derived at mean-field level from a Hamiltonian of the form shown, although a phase diagram of this sort --- with a ``vestigial nematic'' phase --- can arise from fluctuation effects, as discussed below and in Refs.~\cite{goluboviic} and \cite{nie}.  To obtain this sort of phase diagram in mean-field theory, we need to add in an additional nematic order parameter, ${\cal N}$, which transforms as $B_{1g}$ ($x^2-y^2$) under the point group symmetry of the tetragonal crystal.  To the same order as we have considered so far, this involves
\begin{align}
V \to V+&\tilde \alpha {\cal N}+\frac {\mu_{\mathcal{N}}} 2 {\cal N}^2 +\frac{\tilde w} 3 {\cal N}^3+  \frac {1}4 {\cal N}^4 \nonumber\\
+ &\frac \lambda 2 {\cal N} \left[|\phi_1|^2-|\phi_2|^2\right]
\label{nematic}
\end{align}
where the nematic field has been normalized in such a way that the quartic coupling is equal to 1, and where $\tilde \alpha$ and $\tilde w$ both vanish by symmetry in the absence of strain. 
To obtain the phase diagram in Fig.~\ref{fig:MFPhaseDiagrams}f, we have assumed that $\mu_{\mathcal{N}}$ changes sign at a temperature slightly above the putative bicritical point, and  in projecting the phase diagram onto the $\mu-\tilde \mu$ plane we have taken $\mu_{\mathcal{N}} = -\mu_{\mathcal{N}}^* +\mu_{\mathcal{N}}' \mu$, with $\mu_{\mathcal{N}}^*$ small and positive. Coefficients of terms odd in $\cal N$ vary linearly with $\tilde \mu$: $\tilde \alpha = \tilde \alpha'\tilde \mu$ and $\tilde w = \tilde w' \tilde \mu$.  Now there is a nematic critical point at $\tilde \mu=0$ and $\mu=\mu^*\equiv \mu_{\mathcal{N}}^*/\mu_{\mathcal{N}}'$,  while the two critical endpoints (where the CDW ordering temperatures intersect a first-order line) occur at a smaller value of $\mu$  which (for example) is at $\mu=\mu^*\left[ \lambda \mu_{\mathcal{N}}/(1+\lambda \mu_{\mathcal{N}}')\right]$ in the limit $\tilde \mu \to 0$.

\end{itemize}

\subsection{Emergent {$\mathbb{Z}_2$} symmetry: the orthorhombic case}\label{subsec:orthomeanfield}

\begin{figure*}
    \centering
    \includegraphics[width=\linewidth]{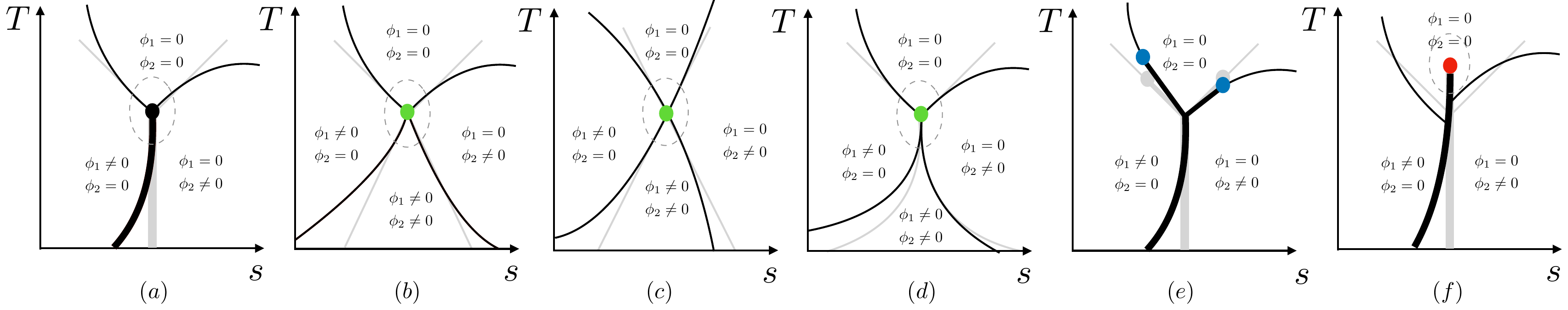}
    \caption{Mean-field phase diagrams in the non-$\mathbb{Z}_2$-symmetric (orthorhombic) cases discussed in Sec.~\ref{subsec:orthomeanfield}: (a) bicritical point, (b) tetracritical point, (c)  decoupled tetracritical point,  (d) $O(4)$-symmetric tetracritical point, (e) tricritical points, and (f) vestigial metanematic. Thin and thick black lines denote, respectively, continuous and first-order transitions.  Dashed ovals indicate regimes of approximate emergent $\mathbb{Z}_2$ symmetry.  Gray lines are the corresponding phase boundaries from Fig.~\ref{fig:MFPhaseDiagrams}.}
    \label{fig:orthoMFPhaseDiagrams}
\end{figure*}

We now consider modifications to the mean-field phase diagrams that result from considering a system which is weakly orthorhombic, i.e. where $\tilde H \neq 0$ even for $s=0$. Since at mean-field level  we can neglect the gradient terms, to quartic order there are only two terms that break the $\mathbb{Z}_2$ symmetry under exchange  $\phi_1 \leftrightarrow \phi_2$: the quadratic term, $\tilde \mu$, and the quartic term, $\tilde u$.  Rescaling the order parameter fields such that
$\phi_1 \to Z \phi_1$ and $\phi_2 \to Z^{-1} \phi_2$ with $Z=\left[(u+\tilde u)/(u-\tilde u)\right]^{-1/8}$
preserves the form of the potential $V+\tilde V$, but with shifted parameters:
\begin{align}
&\mu\to\left[(Z^4+1)\mu+ (Z^4-1)\tilde \mu\right]/({2Z^2}),\nonumber \\
&\tilde \mu\to\left[(Z^4-1)\mu+ (Z^4+1)\tilde \mu\right]/({2Z^2}),\nonumber \\
&u\to \sqrt{u^2-\tilde u^2},\\
&\gamma \to \gamma+u -\sqrt{u^2-\tilde u^2}, \nonumber \\
&\tilde u \to 0. \nonumber
\end{align}

We see therefore that (to the extent that we can neglect higher order terms in powers of $\phi_j$) the theory possesses an emergent $\mathbb{Z}_2$ symmetry under the exchange of $\phi_1$ and $\phi_2$ and $\tilde \mu \to -\tilde \mu$.  In this sense, it behaves as if it were tetragonal whether it is or not.   Of course, the fact that this is not a true symmetry will be reflected in the presence of higher order terms, such as $\tilde u_6\left[|\phi_1|^2+|\phi_2|^2\right]\left[|\phi_1|^4-|\phi_2|^4\right]$, which do not vanish under the rescaling transformation.  Consequently, while an appropriate coordinate transformation in the $T-s$ plane will make the phase diagram of the orthorhombic system look like that of a tetragonal system close enough to criticality that higher order terms in $V+\tilde V$ can be ignored, further from criticality, where the higher order terms begin to play a role, the emergent $\mathbb{Z}_2$ symmetry will be increasingly violated, as reflected in the schematic phase diagrams in Fig.~\ref{fig:orthoMFPhaseDiagrams}.

The situation is somewhat subtle for the tricritical case in Fig.~\ref{fig:orthoMFPhaseDiagrams}e since higher order (at least sixth) terms play a role in the phase diagram near criticality.  Thus the extent to which this system exhibits an approximate $\mathbb{Z}_2$ symmetry can depend on other assumptions.

The case of the vestigial nematic, Fig.~\ref{fig:orthoMFPhaseDiagrams}f, needs to be handled separately.  In the first place there is not, strictly speaking, a nematic phase defined by a spontaneously broken symmetry, since the symmetry in question is explicitly broken.  This is reflected in the presence of odd terms, $\tilde \alpha$ and $\tilde w$ in the  effective potential in Eq.~\eqref{nematic}, 
where now no symmetry requires that $\tilde \mu$, $\tilde \alpha$, and $\tilde w$ all vanish simultaneously.
Thus the first-order line immediately below the critical point corresponds to a line of metanematic transitions (where the value of ${\cal N}$ jumps discontinuously --- typically from a negative to a positive value).  The critical point at the end --- what was formerly the nematic critical point --- is now analogous to the critical point that terminates the liquid-gas coexistence line in the phase diagram of water or other common liquids. 

That the vicinity of the critical point possesses an emergent $\mathbb{Z}_2$ symmetry can be seen by making a shift of the order parameter, ${\cal N} \to {\cal N}+\bar{{\cal N}}$, where $\bar{{\cal N}}=-
\tilde w/3 
$ is chosen to cancel the third order term  --- this is again analogous to Landau theory for the liquid-gas transition~\cite{chaikin_lubensky_1995}.  Then, with proper rescaling of ${\cal N}$, this replaces Eq.~\eqref{nematic} 
with an expression of the same form, but with $\tilde w=0$. The metanematic critical point at $\mu=\mu^*$ and $\tilde \mu = \tilde \mu^*\equiv -\alpha^*/\alpha'$ thus has the same form as in the tetragonal case.  However, as shown in the figure, the critical end points (where the CDW transitions end on the metanematic line) are split by order $\lambda \bar{{\cal N}}$, thus spoiling the $\mathbb{Z}_2$ symmetry as one moves away from the critical point.

\section{Fluctuation Effects}\label{sec:fluctuations}

Mean-field theory tends to work remarkably well in metallic systems --- presumably because the effective range of interactions is relatively long.  The coupling of the nematic component of the order parameter to the strain field tends to further reduce fluctuations effects (see Ref.~\cite{garst} and references therein).  It is, nonetheless, interesting in a more general context to analyze what of the above structure survives (or may even be enhanced by)  fluctuation effects in the true asymptotic critical regime.

The most basic question to be addressed is the stability of the various mean-field phase diagrams to fluctuations.  The issue concerning the stability of $O(N)\times O(N)$ (or more generally $O(N)\times O(M)$) multicritical points has been addressed in several different ways, notably through the use of the $\epsilon$ expansion~\footnote{
For recent overviews of the 
$\epsilon$ expansion  for multicritical points, see Refs.~\cite{aharony3, AharonyFisher}}, the functional renormalization group~\cite{frg}, and, more recently, using conformal bootstrap to address the problem directly in three dimensions~\cite{bootstrapreview, bootstrapsnowmass}. 
Generalizing 
the $O(2)^2$ symmetries discussed 
so far to $O(N)\times O(M)$ 
would 
make no qualitative difference to our mean-field analysis, but is useful in discussing 
effects of fluctuations; 
however we will also return to the $N=M=2$ case which is of particular interest.

In the context of the $\epsilon$ expansion, there are three fixed points (in addition to the always unstable Gaussian fixed point) which correspond to the decoupled tetracritical point, a higher symmetry $O(2N)$ symmetric multicritical point, and an interacting ``biconical'' fixed point.  However, for given $N$, only one of these is ever stable with the other two being unstable --- i.e. the other two require at least one additional parameter to be fine-tuned (in addition to the usual two associated with  ``stable'' multicritical points).  

A number of conclusions can been reached based on nonperturbative lines of analysis~\cite{Henriksson}. 
For $N=1$, the $O(2)$ symmetric fixed point, familiar from studies of the Ashkin-Teller model, is stable~\footnote{Note that, throughout this section, we discuss $O(2)$ order parameters interchangeably with those which we have previously called $U(1) \equiv SO(2)$ ones. The distinction between the two groups is of no consequence for our purposes. }.
This is one of the most remarkable examples of an emergent symmetry.  However,  it is generally accepted that the $O(2N)$ symmetric fixed point is (at least weakly) unstable for all $N>1$~\cite{O3CubicInstability,calabrese}. 

For $N=2$ the situation is somewhat  subtle. 
Certainly, the decoupled fixed point is stable for $N\geq 2$. As viewed from the perspective of the $\epsilon$ expansion, this suggests that this is the only stable multicritical point, which among other things would imply that there is no stable $O(N)\times O(N)$ bicritical point with $N\geq 2$.  Recently, however, initial evidence has emerged from conformal bootstrap of the existence of an additional stable conformal field theory that might correspond to the long-sought bicritical point.  We will return to this below.

We now comment on the tricritical scenario. From the fact that there is no stable bicritical fixed point in the context of the $\epsilon$ expansion, it was suggested by Aharony \textit{et al.}~\cite{Aharony2,aharony3, AharonyFisher} that it was likely to be fluctuation-driven first order.  This further led to the suggestion that a mean-field bicritical phase diagram should instead exhibit two tricritical points, and thus be of the form shown in Figs.~\ref{fig:MFPhaseDiagrams}e and \ref{fig:orthoMFPhaseDiagrams}e.

The tricritical points in this case are readily characterized since $d=3$ is the upper critical dimension, and so they should exhibit mean-field behavior up to logarithmic corrections.
Clearly, each tricritical point involves the ordering of only one of the CDW components, so there can be no question of an emergent $\mathbb{Z}_2$ symmetry.  

Another possible form of a phase diagram with a fluctuation driven first-order transition  is that corresponding to a vestigial nematic phase, as in Figs.~\ref{fig:MFPhaseDiagrams}f and \ref{fig:orthoMFPhaseDiagrams}f.   It has been shown~\cite{nie} that such a vestigial nematic phase arises from fluctuations in a sufficiently layered (quasi-2D) system with $N>2$, and strong arguments suggest that this is true as well for $N=2$.  There thus appear to be circumstances in which this form of a phase diagram arises organically from CDW fluctuations, without need to introduce an explicit nematic order parameter.  (See also  Refs. \cite{AeppliEmeryMe} and \cite{goluboviic}.)

In this case the critical point in question is in the 3D Ising universality class.  For the tetragonal case, this is obvious in that there is an Ising symmetry --- which corresponds to the $\mathbb{Z}_2$ symmetry under exchange $\phi_1 \leftrightarrow \phi_2$ --- that is broken below the critical point.  But in the orthorhombic case, the analogy with the liquid-gas critical point is more apt, in that there is no actual broken symmetry associated with the first-order line below the critical point.  However, the present discussion puts a somewhat different perspective on this familiar problem --- there is an emergent $\mathbb{Z}_2$ symmetry at the critical point which is broken along the first-order line below the critical point.  Correspondingly, there is presumably a Widom line, at which the emergent $\mathbb{Z}_2$ symmetry is best defined, that extends above the critical point, and which should be observable as a peak in the nematic susceptibility~\cite{Z2experiment}.

The conformal bootstrap offers a new approach to identifying possible critical phenomena --- especially in 3D~\cite{bootstrapreview, bootstrapsnowmass}.  There is not yet  any systematic classification of all possible 3D conformal field theories, but (in many cases) where such field theories have been identified and characterized from this approach, these provide important complements to the $\epsilon$ expansion and other more familiar approaches to critical phenomena.  
For instance,  the perturbative stability of the decoupled $O(2)\times O(2)$ tetracritical point can  be corroborated  by knowing precisely the critical exponents of the $O(2)$ critical point~\cite{Henriksson}.

In this context, it is interesting to note that there is preliminary indication of the existence of at least one additional stable theory with $O(2)\times O(2)$ symmetry, distinct from the decoupled theory~\cite{stergiou1,stergiou2, stergiou3}.  The results here are not definitive, and the identification of the physical meaning of this additional critical theory --- if it indeed exists --- is also not established.  It is tempting, however, to identify this as the long-sought bicritical theory.  In this context, it is interesting to note that the theories bootstrapped in these works actually have a higher, $(O(2)\times O(2))\rtimes \mathbb{Z}_2$ symmetry, and so would correspond to a critical point with  emergent $\mathbb{Z}_2$ symmetry under exchange of the two order parameters.

Finally, there is  an issue concerning the presence or absence of an emergent $SO(3)$ spatial rotational symmetry at criticality.  This is a feature of 
an $O(2)$-symmetric critical point, such as occurs along the narrow solid phase boundaries in the figures. 
In particular, by appropriate rescaling of the length scales in the $x$, $y$, and $z$ directions, the system at criticality --- even on an underlying orthorhombic lattice --- can be described by an $SO(3)$ rotationally symmetric effective field theory. The same analysis can be applied to the tricritical points or to the nematic or metanematic critical points, such as those illustrated in Figs.~\ref{fig:MFPhaseDiagrams}f and \ref{fig:orthoMFPhaseDiagrams}f.  It would presumably be true at a  bicritical point of the sort shown in Figs.~\ref{fig:MFPhaseDiagrams}e and \ref{fig:orthoMFPhaseDiagrams}e, if such a critical point indeed arises.

In contrast, it is easy to see that rotational symmetry does not emerge  at the decoupled tetracritical points, because of  the presence of an effective ``spin-orbit coupling'' that reflects the fact that the broken symmetries involved are spatial symmetries.  Specifically,  even in the tetragonal case, while the  effective field theory for $\phi_1$ can be made rotationally symmetric by rescaling $x$, $y$, and $z$  appropriately,  to achieve the same result  for $\phi_2$ requires interchanging the scale factors for $x$ and $y$.
In addition, if the system is orthorhombic, there is a different scale required  for the magnitude of fluctuations of $\phi_1$ and $\phi_2$ and for the $z$ coordinate that enters the two theories.  Thus, even at criticality, there is only $C_2$ rotational symmetry for the orthorhombic system, while for the tetragonal case the  $C_4$ rotational symmetry  is represented as a  further symmetry under $ \phi_1 \leftrightarrow \phi_2$ and $ x \leftrightarrow  y$.

\section{Discussion}\label{sec:discussion}

The analysis in the present paper was motivated by the observation of an apparent bicritical phase diagram with an emergent tetragonal symmetry (signaled by a strong peak in the elastoresistivity near said multicritical point) in ErTe$_3$, a weakly orthorhombic, quasi-two-dimensional  (layered) material with unidirectional CDW order whose direction can be reoriented with the application of uniaxial stress.  Further studies of this material with the  goal of exploring the behavior closer to the putative bicritical point would certainly be interesting.  In particular, many of the theoretically most interesting issues concern fluctuation effects not captured by mean-field theory, while at present it is not clear whether the experiments approach close enough to criticality to be sensitive to such effects.

It is worth noting that similar considerations apply to systems with unidirectional spin density waves as well, with the modification that in that case the order parameter symmetries associated with the two order parameter fields are richer than in the present case.  For instance, for a commensurate unidirectional collinear antiferromagnet --- i.e. the $(0,\pi)-(\pi,0)$ ``stripe'' antiferromagnet seen in the Fe-based superconductors~\cite{schmalianreview} --- the two order parameter fields (to the extent that spin-orbit coupling is negligible) are three-component fields, so the relevant symmetry is $O(3)\times O(3)$ with, for tetragonal systems, an additional $\mathbb{Z}_2$ under exchange.  Many other examples exist, with other symmetries or near symmetries, including materials with still lower point-group symmetry,  in various strongly interacting electronic materials.

The analysis --- with suitable modifications --- is also of possible relevance to unconventional superconductors under circumstances in which  by tuning some non-thermal parameter, e.g. pressure or alloy concentrations, the system can be tuned from a regime in which there is one form of superconducting order to a regime in which there is another.  Here, again, one expects a multicritical phase diagram at the point at which the two forms of superconductivity have the same $T_c$.  An example of this is an apparent change from an extended s-wave to a d-wave superconductor that appears as a function of chemical substitution in Ba$_{1-x}$K$_x$Fe$_2$As$_2$~\cite{egor}.

From a theory perspective, it remains to determine whether or not a stable bicritical point not accessible from the $\epsilon$ expansion exists, what are the best experimental tests of the existence of an emergent $\mathbb{Z}_2$ symmetry in the critical regime, and more generally how to nail down the topology of the phase diagram in the asymptotic multicritical regime where mean-field considerations are insufficient.
\bigskip

\acknowledgments

We gratefully acknowledge useful discussions and correspondence with Vladimir Calvera, Eduardo Fradkin, Andreas Stergiou and Gilles Tarjus.  
The work was supported by the Department of Energy, Office of Basic Energy Sciences, under Contract No. DE-AC02-76SF00515. 
A.P. is supported in part by a Stanford  Graduate Fellowship. 
A.G.S. was supported in part by the National Science Foundation Graduate Research Fellowship Program under Grant No. DGE-1656518 and a DOE SCGSR award. 
S.A.K. was further supported by a Leverhulme Trust International Professorship Grant No. LIP-202-014 at Oxford.

\bibliography{EmergentZ2.bib}

\end{document}